\documentclass[floats,floatfix,twocolumn,superscriptaddress,nofootinbib,showpacs]{revtex4-1}

\pdfoutput=1
\usepackage{amsfonts,amssymb,amsmath}
\usepackage[usenames,dvipsnames]{xcolor}
\definecolor{292}{rgb}{0.3843, 0.6588, 0.8980}
\usepackage{scrextend}
\usepackage{tablefootnote}
\usepackage[unicode, colorlinks=true, 
linkcolor=blue, 
urlcolor=blue, 
citecolor=292]{hyperref}
\usepackage[all]{hypcap}
\usepackage{graphicx}
\usepackage{xspace}
\usepackage{tikz}
\usepackage{float}
\usetikzlibrary{shapes,arrows}
\usetikzlibrary{matrix}
\usetikzlibrary{shapes.multipart}
\usetikzlibrary{er,positioning}
\usepackage[normalem]{ulem} %for \sout
\usepackage{url}
\usepackage{color,soul}
\usepackage[caption=false]{subfig}
\usepackage[T1]{fontenc}
\usepackage{lmodern}
\usepackage[utf8]{inputenc}
\usepackage{microtype}
\usepackage{array}
\newcolumntype{C}[1]{>{\centering\let\newline\\\arraybackslash\hspace{0pt}}m{#1}}
\usepackage{makecell}
\usepackage{diagbox}
\usepackage{orcidlink}

\newcommand{\Caltech}{\affiliation{Theoretical Astrophysics,
		Walter Burke Institute for Theoretical Physics,\\
		California Institute of Technology, Pasadena, California 91125, USA}}
\newcommand{\Cornell}{\affiliation{Cornell Center for Astrophysics and
		Planetary Science, Cornell University, Ithaca, New York 14853, USA}}
\newcommand{\PennState}{\affiliation{Institute for Gravitation and the Cosmos and Physics Department, Penn State, University Park, Pennsylvania 16802, USA}}
\newcommand{\MaxPlanck}{\affiliation{Max Planck Institute for Gravitational Physics (Albert Einstein Institute), Am M{\"u}hlenberg 1, Potsdam 14476, Germany}}

\makeatletter
\newsavebox\myboxA
\newsavebox\myboxB
\newlength\mylenA

\newcommand*\xoverline[2][0.75]{%
	\sbox{\myboxA}{$\m@th#2$}%
	\setbox\myboxB\null% Phantom box
	\ht\myboxB=\ht\myboxA%
	\dp\myboxB=\dp\myboxA%
	\wd\myboxB=#1\wd\myboxA% Scale phantom
	\sbox\myboxB{$\m@th\overline{\copy\myboxB}$}%  Overlined phantom
	\setlength\mylenA{\the\wd\myboxA}%   calc width diff
	\addtolength\mylenA{-\the\wd\myboxB}%
	\ifdim\wd\myboxB<\wd\myboxA%
	\rlap{\hskip 0.5\mylenA\usebox\myboxB}{\usebox\myboxA}%
	\else
	\hskip -0.5\mylenA\rlap{\usebox\myboxA}{\hskip 0.5\mylenA\usebox\myboxB}%
	\fi}
\makeatother

\newcommand{\cmmnt}[1]{\ignorespaces}

\begin{document}
	
\title{Adding Gravitational Memory to Waveform Catalogs using BMS Balance Laws}
	
%%% Tentative Author Order %%%
\author{Keefe Mitman
\orcidlink{0000-0003-0276-3856}}
\email{kmitman@caltech.edu}
\Caltech
\author{Dante A. B. Iozzo
\orcidlink{0000-0002-7244-1900}}
\Cornell
\author{Neev Khera}
\PennState
% Alphabetical Ordering
\author{Michael Boyle
\orcidlink{0000-0002-5075-5116}}
\Cornell
\author{\\Tommaso De Lorenzo}
\PennState
\author{Nils Deppe
\orcidlink{0000-0003-4557-4115}}
\Caltech
\author{Lawrence E. Kidder
\orcidlink{0000-0001-5392-7342}}
\Cornell
\author{Jordan Moxon
\orcidlink{0000-0001-9891-8677}}
\Caltech
\author{\\Harald P. Pfeiffer
\orcidlink{0000-0001-9288-519X}}
\MaxPlanck
\author{Mark A. Scheel
\orcidlink{0000-0001-6656-9134}}
\Caltech
\author{Saul A. Teukolsky
\orcidlink{0000-0001-9765-4526}}
\Caltech
\Cornell
\author{William Throwe
\orcidlink{0000-0001-5059-4378}}
\Cornell
	
\date{\today}
	
\begin{abstract}
	\noindent Accurate models of gravitational waves from merging binary black holes are crucial for detectors to measure events and extract new science. One important feature that is currently missing from the Simulating eXtreme Spacetimes (SXS) Collaboration's catalog of waveforms for merging black holes, and other waveform catalogs, is the gravitational memory effect: a persistent, physical change to spacetime that is induced by the passage of transient radiation. We find, however, that by exploiting the Bondi-Metzner-Sachs (BMS) balance laws, which come from the extended BMS transformations, we can correct the strain waveforms in the SXS catalog to include the missing displacement memory. Our results show that these corrected waveforms satisfy the BMS balance laws to a much higher degree of accuracy. Furthermore, we find that these corrected strain waveforms coincide especially well with the waveforms obtained from Cauchy-characteristic extraction (CCE) that already exhibit memory effects. These corrected strain waveforms also evade the transient junk effects that are currently present in CCE waveforms. Last, we make our code for computing these contributions to the BMS balance laws and memory publicly available as a part of the python package $\texttt{sxs}$, thus enabling anyone to evaluate the expected memory effects and violation of the BMS balance laws.
\end{abstract}
	
\maketitle
	
%%%%%%%%%%%%%%%%%%%%%%%%%%%%%%%%%%%%%%%%%%%%%%%%%%%%%%%%%%%
\section{Introduction}
%%%%%%%%%%%%%%%%%%%%%%%%%%%%%%%%%%%%%%%%%%%%%%%%%%%%%%%%%%%
\label{sec:introduction}
	
When Bondi, van der Burg, Metzner, and Sachs (BMS) tried to recover the Poincar\'e group of special relativity as the symmetry group of asymptotically flat spacetimes in general relativity, they instead found an unexpected infinite-dimensional group of transformations, known as the \emph{BMS group} \cite{Bondi,Sachs}. Fundamentally, the BMS group extends the Poincar\'e group with an infinite number of transformations called \emph{supertranslations}.\footnote{Formally, the BMS group is simply a semidirect product of the Lorentz group with this infinite-dimensional Abelian group of spacetime supertranslations, which contains the usual translations as a normal subgroup. If one represents spacetime translations as the $\ell<2$ spherical harmonics, then supertranslations can be viewed as the $\ell\geq2$ spherical harmonics, i.e., when acted on by a supertranslation the Bondi time $u\equiv t-r$ changes as
\begin{align}
u'=u-\alpha(\theta,\phi)\quad\text{for}\quad\alpha=\sum\limits_{\ell\geq2}\sum\limits_{m\leq|\ell|}\alpha_{\ell m}Y_{\ell m}(\theta,\phi)
\end{align}
with $\alpha_{\ell,m}=(-1)^{m}\bar{\alpha}_{\ell,-m}$ to ensure that $u'$ is real.} More recent research~\cite{de_Boer_2003, banks2003critique, Barnich_2010, barnich2011supertranslations, Barnich_2011, Kapec_2014, Kapec_2017, He_2017} motivates the consideration of an extended BMS group that includes another set of transformations known as \emph{super-Lorentz transformations}.\footnote{Originally, these transformations were known as superrotations. They can be thought of as a Virasoro-like symmetry acting on the sphere at asymptotic infinity, i.e., the $|m|\geq2$ elements of the Virasoro algebra, which is just an extension of the more common group of M\"obius transformations.} When these transformations are included, the group is then called the \emph{extended} BMS group. Like rotations and boosts in special relativity, we refer to the magnetic parity piece of super-Lorentz transformations as \emph{superrotations} and the electric parity piece as \emph{superboosts}.

One of the extended BMS group's more useful features is that for each transformation there is a corresponding balance law. Just as the translation symmetries lead to the four-momentum and its balance laws at null infinity, the supertranslations and super-Lorentz transformations of the extended BMS group induce ``super'' balance laws. These super,  or just BMS, balance laws can be extracted from the Einstein field equations by examining certain evolution equations~\cite{Barnich_2011, Barnich_2012, Flanagan_2017, Comp_re_2018}. There is an infinite tower of balance laws: one for each point on the two-sphere or, equivalently, one for each spherical harmonic mode. Furthermore, the BMS flux part of these balance laws can be broken into two contributions, called ``hard'' and ``soft,'' which are based on the order in which the gravitational wave strain appears: nonlinear for the hard contribution and linear for the curious soft contribution. An example of the relationship between BMS charges and BMS fluxes is the well-known mass loss equation~\cite{Bondi}
\begin{align}
\label{eq:example}
\dot{m} = -\frac{1}{4}\dot{h}\dot{\bar{h}} + \frac{1}{4}\text{Re}\left[\eth^{2}\dot{h}\right],
\end{align}
where $m$ is the Bondi mass aspect, $h$ is the strain, and $\eth$ is the spin-weight operator (see Sec.~\ref{sec:conventions} for more details). In Eq.~\eqref{eq:example}, the left-hand side is the BMS charge, while the right-hand side is the BMS flux, with the ``$\dot{h}\dot{\bar{h}}$'' term being the hard contribution (notice that this term is just the time derivative of the energy flux) and the ``$\eth^{2}\dot{h}$'' term being the soft contribution.\footnote{Originally~\cite{Bondi}, Eq.~\eqref{eq:example} was written after it had been integrated over the two-sphere so the ``$\eth^{2}\dot{h}$'' term could be ignored since its $\ell=0,1$ modes are zero.}
When integrated with respect to time, the soft contribution to the total BMS flux is then proportional to the \emph{gravitational memory}, or just the \emph{memory}, of the gravitational wave. Consequently, the BMS balance laws can be viewed as relating the memory to contributions coming from the BMS charges and the hard part of the flux of an arbitrary system.
	
The gravitational memory effect, on its own, has been studied for decades~\cite{Zeldovich_1974, Thorne_1987, Christodoulou_1991, Thorne_1992, Favata2009a, Favata2009b, Favata_2009_PN, Favata2010, Favata2011}. Only recently~\cite{Strominger_2014, Pasterski_2016, Nichols_2018, Comp_re_2020}, however, have new memory effects started to be understood in relation to the elements of the extended BMS group. Because of this connection, memory has been categorized into three types based on which BMS transformations they correspond to. We outline these types in Table~\ref{tab:memorytypes}, along with their physical meanings.
\begin{table*}[t]
	\label{tab:memorytypes}
	\centering
	\caption{The three types of memory and their physical interpretations.
		%\small
	}
	\begin{tabular}{C{4cm}C{6cm}C{6cm}}
		\Xhline{3\arrayrulewidth}
		Memory type & BMS transformation & Physical interpretation \\
		\colrule
		Displacement memory \phantom{takeupsomespacetoverticallycenter} & Supertranslations \phantom{takeupsomespacetoverticallycenter} & Change in the relative position of two freely falling, initially comoving observers. Appears in the strain. \\
		\colrule
		Spin memory \phantom{takeupsomespacetoverticallycenter} \phantom{takeupsomespacetoverticallycenter} & Superrotations \phantom{takeupsomespacetoverticallycenter} \phantom{takeupsomespacetoverticallycenter} & Change in the relative time delay of two freely falling observers on initially counterorbiting trajectories. Appears in the retarded time integral of the strain. \\
		\colrule
		Center-of-mass memory \phantom{takeupsomespacetoverticallycenter}  \phantom{takeupsomespacetoverticallycenter}  & Superboosts \phantom{takeupsomespacetoverticallycenter} \phantom{takeupsomespacetoverticallycenter} & Change in the relative time delay of two freely falling observers on initially antiparallel trajectories. Appears in the retarded time integral of the strain. \\
		\Xhline{3\arrayrulewidth}
	\end{tabular}
\end{table*}
Apart from these characterizations, memory is also classified by its \emph{ordinary} and \emph{null} parts, with the ordinary part coming from the BMS charges and the null part coming from the hard part of the flux. Besides furthering the physical understanding of gravitational memory, however, the BMS balance laws can also offer practical uses to waveform modeling efforts. As was proposed in~\cite{ashtekar2019compact}, and as we will show throughout this paper, the BMS balance laws can be used to both check the accuracy of waveforms and also improve them if they are missing certain flux contributions.

The framework provided by the extended BMS group is applicable only to asymptotic Bondi gauge waveforms. The waveforms that are extracted at finite radii in current numerical simulations do not fall into this category, however, since the simulation coordinates are not necessarily in the Bondi gauge and there are subleading terms in the waveforms in addition to the asymptotic contribution~\cite{Boyle_2009}. Therefore, these finite-radius waveforms must be either extrapolated~\cite{Iozzo2020} or evolved~\cite{moxon2020improved,moxon2021,CodeSpECTRE} to asymptotic infinity with respect to Bondi coordinates.

Additionally, the extrapolated strain waveforms that are made publicly available in the SXS catalog~\cite{SXSCatalog, Boyle_2019} and other waveform catalogs~\cite{Jani_2016,Healy_2017} fail to capture the displacement memory for an unknown reason~\cite{Favata_2009_PN, mitman2020computation}. But, by using an alternative waveform extraction technique called Cauchy-characteristic extraction (CCE)~\cite{moxon2020improved}, the memory can be captured and the resulting waveforms appear to minimally violate the BMS balance laws~\cite{mitman2020computation}. Our goal in this paper is to use the balance laws to help improve the extrapolated strain waveforms; in particular, to make them exhibit the displacement memory effect so that they are more on a par with the CCE waveforms. Consequently, we not only need to know the total memory, but also how the memory evolves. That is, we must use the finite-time version of the balance laws as opposed to the global version, i.e., we must compare the BMS charges to the BMS fluxes as functions of time rather than the net changes in the two. This version of the BMS balance laws can be rewritten to give the gravitational wave strain $h$ as a combination of two unique terms: an energy flux term $J_{\mathcal{E}}$ and a more oscillatory term $J_{\Psi}$ that depends on the Weyl scalar $\Psi_2$.\footnote{While $J_{\Psi}$ is certainly more oscillatory than $J_{\mathcal{E}}$, it is important to note that $J_{\mathcal E}$ does not always change monotonically and can exhibit oscillatory behavior as well for certain BBH systems, such as precessing systems.} With these new terms, we can then write the balance laws as
\begin{align}
\label{eq:h_balance}
h = J_{\mathcal{E}}+J_{\Psi}.
\end{align}

The energy flux term, which is the primary contribution to the strain's total displacement memory~\cite{mitman2020computation, Talbot_2018}, was first computed in \cite{Talbot_2018} and depends only on the news $\dot{h}$.\footnote{This term was also identified in~\cite{mitman2020computation} as being the null component of the electric parity piece of the displacement memory; see Table~\ref{tab:memorytypes}.} Because the current extrapolated strain waveforms in the SXS catalog fail to capture the displacement memory, we can therefore ask if these waveforms are reasonably modeled by just the $J_{\Psi}$ term. If so, we can then add the $J_{\mathcal{E}}$ term to the extrapolated strain waveforms to see if the violation of the BMS balance laws and the mismatch between extrapolated and CCE waveforms are improved. Indeed, we find empirically that extrapolation gives the $J_{\Psi}$ term to reasonable accuracy and adding the energy flux $J_{\mathcal{E}}$ term gives a more complete and improved waveform.

We find that when the expected memory $J_{\mathcal E}$ is added to the extrapolated strain waveforms, the extrapolated waveforms obey the BMS balance laws to roughly the same degree as those obtained by CCE. Furthermore, we observe that the mismatch between extrapolated and CCE waveforms near merger is considerably improved. We find that the next-leading cause for the violation of the BMS balance laws is due to the part of the waveform that is the primary contribution to the spin memory effect. For an unknown reason, there is already an incomplete\footnote{We find that the spin memory present in the extrapolated strain is roughly 50\% of what is seen in the CCE strain.} spin memory contribution to the extrapolated waveform, unlike the displacement memory contribution which is completely absent~\cite{mitman2020computation}. Unfortunately, this means that we have no justified way to correct the spin memory contribution without using the Weyl scalar $\Psi_{2}$, which has not yet been extracted for the simulations in the SXS catalog. Nonetheless, we find that the extrapolated waveform is indeed a good approximation to the $J_{\Psi}$ term, although it does not wholly capture the spin memory. Therefore, whenever extrapolated waveforms are used for any kind of analysis, they should first be corrected by the $J_{\mathcal{E}}$ term to include memory and thus remain consistent with expectations from theory.

Apart from improving the extrapolated waveforms, we also observe that once the $J_{\mathcal{E}}$ correction is added, the Bondi frame of our extrapolated waveforms after ringdown is nearly the same as that of the waveforms obtained from CCE. This is important because gravitational wave astronomy expects the waveforms produced by numerical relativity to be close to, if not in, the super rest frame\footnote{See Sec.~\ref{sec:BondiFrame} for a more thorough explanation.} to circumvent frame ambiguities. Without the displacement memory correction, however, the extrapolated waveforms in the SXS catalog would prompt one to think that the system was already fairly close to the super rest frame, which would be markedly incorrect. Consequently, the displacement memory correction not only serves as an important improvement to the extrapolated waveforms, but is also crucial for correctly comparing the Bondi frame of the multitude of waveforms in the SXS catalog to the super rest frame and ensuring that waveforms, when being compared to each other, are in the same Bondi frame.

We make our code for computing these contributions to the BMS balance laws and memory publicly available as a part of the python package \texttt{sxs}~\cite{sxs_python}.

%%%%%%%%%%%%%%%%%%%%%%%%%%%%%%%%%%%%%%%%%%%%%%%%%%%%%%%%%%%
\subsection{Overview}
%%%%%%%%%%%%%%%%%%%%%%%%%%%%%%%%%%%%%%%%%%%%%%%%%%%%%%%%%%%
\label{sec:overview}

We organize our computations and results as follows. In Sec.~\ref{sec:memory} we outline the BMS balance laws that we will use to measure waveform accuracy, using results obtained in~\cite{ashtekar2019compact} and~\cite{mitman2020computation}. Following this, in Sec.~\ref{sec:extrap} we illustrate the main sources of BMS balance law violations for the extrapolated waveforms in comparison to CCE and then explain which of these sources we can correct and why. Next, in Sec.~\ref{sec:extrapwithmemory} we present the changes to the extrapolated waveforms once displacement memory is included, explore other sources of the violation of the BMS balance laws, and then compare the corrected extrapolated waveforms to the waveforms that are obtained using CCE. Finally, in Sec.~\ref{sec:BondiFrame} we measure the Bondi frame of the corrected extrapolated waveforms and discuss prospects for mapping to the super rest frame.

%%%%%%%%%%%%%%%%%%%%%%%%%%%%%%%%%%%%%%%%%%%%%%%%%%%%%%%%%%%
\subsection{Conventions}
%%%%%%%%%%%%%%%%%%%%%%%%%%%%%%%%%%%%%%%%%%%%%%%%%%%%%%%%%%%
\label{sec:conventions}

We set $c=G=1$ and use the Newman-Penrose (NP) convention for the spin-weight operator $\eth$ \cite{doi:10.1063/1.1724257} to coincide with the recent work of~\cite{moxon2020improved} and~\cite{mitman2020computation},
\begin{align}
\eth_{s}Y_{\ell m}&=+\sqrt{(\ell-s)(\ell+s+1)}_{s+1}Y_{\ell m},\\
\bar{\eth}_{s}Y_{\ell m}&=-\sqrt{(\ell+s)(\ell-s+1)}_{s-1}Y_{\ell m}
\end{align}
We denote the strain\footnote{We explicitly define the strain as described in Appendix C of~\cite{Boyle_2019}.} by $h$ and use the variable $J$ to represent any contribution to the strain that comes from the BMS balance laws that we compute with our various numerical outputs. Further, we represent the strain in a spin-weight $-2$ spherical harmonic basis,
\begin{align}
h(u,\theta,\phi)=\sum\limits_{\ell,m}h_{\ell m}(u)\,{}_{-2}Y_{\ell m}(\theta,\phi),
\end{align}
where $u\equiv t-r$ is the Bondi time. For our calculations, we use the operators $D^{2}$ and $\mathfrak{D}$, which we construct as
\begin{subequations}
	\begin{align}
	\label{eq:Laplacian}
	D^{2}&=\bar{\eth}\eth,\\
	\label{eq:operator}
	\mathfrak{D}&=\frac{1}{8}D^2\left(D^2+2\right).
	\end{align}
\end{subequations}
Notice that $D^{2}$ is just the Laplacian on the two-sphere when it acts on spin-weight 0 functions. The actions of these operators on spin-weight 0 functions are
\begin{subequations}
	\begin{align}
	D^{2}Y_{\ell m}&=-\ell(\ell+1)Y_{\ell m},\\
	\mathfrak{D}Y_{\ell m}&=\frac{1}{8}(\ell+2)(\ell+1)\ell(\ell-1)Y_{\ell m}.
	\end{align}
\end{subequations}

%%%%%%%%%%%%%%%%%%%%%%%%%%%%%%%%%%%%%%%%%%%%%%%%%%%%%%%%%%%
\section{BMS Balance Laws and Memory}
%%%%%%%%%%%%%%%%%%%%%%%%%%%%%%%%%%%%%%%%%%%%%%%%%%%%%%%%%%%
\label{sec:memory}

We now review some of the work on BMS balance laws and present equations that will prove useful to our analysis. As described in~\cite{ashtekar2019compact}, the supermomentum balance law, which gives the electric parity part of the strain, can be written rather nicely as
\begin{align}
\label{eq:eth2h_real}
\text{Re}&\left[\eth^2h\right]=\nonumber\\
&\int_{-\infty}^{u}|\dot{h}|^2\,du-4\text{Re}\left[\Psi_{2}+\frac{1}{4}\dot{h}\bar{h}\right]-4M_{\mathrm{ADM}},
\end{align}
for a system with ADM mass $M_\text{ADM}$, if one uses the ``post-Newtonian Bondi frame'' where $h\rightarrow0$ as $u\rightarrow-\infty$. The magnetic parity part of the strain can be obtained from the relation between $h$ and $\text{Im}[\Psi_{2}]$:
\begin{align} \label{eq:eth2h_imag}
\text{Im}\left[\eth^2h\right]=-4\text{Im}\left[\Psi_{2}+\frac{1}{4}\dot{h}\bar{h}\right].
\end{align}
Consequently, by combining Eqs.~\eqref{eq:eth2h_real} and~\eqref{eq:eth2h_imag} one has
\begin{align}
\eth^2h=\int_{-\infty}^{u}|\dot{h}|^2\,du-4\left(\Psi_{2}+\frac{1}{4}\dot{h}\bar{h}\right)-4M_{\mathrm{ADM}},
\end{align}
or equivalently
\begin{align}
\label{eq:trueBMS}
h=\frac{1}{2}\bar{\eth}^2\mathfrak{D}^{-1}\left[\frac{1}{4}\int_{-\infty}^{u}|\dot{h}|^2\,du-\left(\Psi_{2}+\frac{1}{4}\dot{h}\bar{h}\right)\right],\footnotemark
\end{align}
\footnotetext[9]{The operator $\tfrac{1}{8}\bar{\eth}^2\mathfrak{D}^{-1}$ is equivalent to $\eth^{-2}$ on spin-weight 0 functions, but is in a form convenient for numerical evaluation.}where $\mathfrak{D}^{-1}$ is defined to map the $\ell\leq1$ modes to zero. Equation~\eqref{eq:trueBMS} represents an infinite tower of balance laws: one for each point on the sphere or, alternatively, one for each spin-weight $-2$ spherical harmonic mode $h_{lm}$.

Numerically, however, we do not have access to the full past, i.e., $u\rightarrow-\infty$. Instead, we are restricted to some finite start time $u_{1}$. This truncation of the time integral in Eq.~\eqref{eq:trueBMS} will cause our computation of the right-hand side to differ from the limit $u_1\to -\infty$ by an unknown angle-dependent constant. Additionally, there is another unknown angle-dependent discrepancy that arises because our numerical waveforms are not necessarily in the post-Newtonian Bondi frame, as required by Eq.~\eqref{eq:trueBMS}. Consequently, when we compute the BMS strain $J$ from the BMS balance laws as
\begin{align}
\label{eq:J}
J\equiv\frac{1}{2}\bar{\eth}^2\mathfrak{D}^{-1}\left[\frac{1}{4}\int_{u_{1}}^{u}|\dot{h}|^2\,du-\left(\Psi_{2}+\frac{1}{4}\dot{h}\bar{h}\right)\right]+\alpha,
\end{align}
we need to account for these angle-dependent shifts by including an unknown constant $\alpha\equiv\alpha(\theta,\phi)$. However, because we have access to both $h$, the asymptotic strain obtained from a numerical spacetime, and $J$, the strain computed from the BMS balance laws in Eq.~\eqref{eq:J}, we may simply solve for $\alpha$ either by minimizing the $L^{2}$ norm of the difference between the two waveforms or by making the waveforms agree on the final time step of the simulation. We may then check the violation of the BMS balance laws in the numerical waveforms via
\begin{align}
\label{eq:hJ}
h-J=0,
\end{align}
where, again, $h$ is the strain output by numerical relativity and $J$ is from Eq.~\eqref{eq:J}, with $\alpha$ being the constant needed to make $h$ and $J$ agree on the simulation's final time step. We choose the final time step, rather than optimizing over all time steps, because this makes our comparison most accurate the farthest away from junk radiation. Equations~\eqref{eq:J} and~\eqref{eq:hJ} are then all that are needed to check the violation of the BMS balance laws. 

As mentioned before, Eq.~\eqref{eq:h_balance} can be used to break $J$ into an energy flux term $J_{\mathcal{E}}$---the primary source of the displacement memory---and a more oscillatory term $J_{\Psi}$,
\begin{subequations}
	\begin{align}
	J_{\mathcal{E}}&=\frac{1}{2}\bar{\eth}^2\mathfrak{D}^{-1}\left[\frac{1}{4}\int_{u_{1}}^{u}|\dot{h}|^2\,du\right]+\alpha,\\
	J_{\Psi}&=\frac{1}{2}\bar{\eth}^2\mathfrak{D}^{-1}\left[-\left(\Psi_{2}+\frac{1}{4}\dot{h}\bar{h}\right)\right].
	\end{align}
\end{subequations}
The hypothesis we test below is that the SXS extrapolated strain waveforms contain only the primarily oscillatory piece $J_{\Psi}$, and one can simply compute and then add the energy flux piece $J_{\mathcal{E}}$ to obtain a waveform with the correct gravitational wave memory. If true, this procedure can then be performed for all of the numerical simulations in the public SXS catalog to substantially improve the extrapolated strain waveforms.

While Eqs.~\eqref{eq:J} and~\eqref{eq:hJ} are all that are needed to check the violation of the BMS balance laws, in~\cite{mitman2020computation} it was shown that $J$ can be decomposed into four terms, which more directly relate to the various memory effects and prove useful when examining numerical waveforms. These terms are
\begin{align}
J=J_{m}+J_{\mathcal{E}}+J_{\widehat{N}}+J_{\mathcal{J}},
\end{align}
where
\begin{subequations}
	\label{eq:Js}
	\begin{align}
	\label{eq:JM}
	J_{m}&=\frac{1}{2}\bar{\eth}^2\mathfrak{D}^{-1}m,\\
	\label{eq:JE}
	J_{\mathcal{E}}&=\frac{1}{2}\bar{\eth}^2\mathfrak{D}^{-1}\left[\frac{1}{4}\int_{u_{1}}^{u}|\dot{h}|^2\,du\right]+\alpha,\\
	\label{eq:JN}
	J_{\widehat{N}}&=\frac{1}{2}i\bar{\eth}^2\mathfrak{D}^{-1}D^{-2}\text{Im}\left[\bar{\eth}\left(\partial_{u}\widehat{N}\right)\right],\\
	\label{eq:JJ}
	J_{\mathcal{J}}&=\frac{1}{2}i\bar{\eth}^2\mathfrak{D}^{-1}D^{-2}\text{Im}\nonumber\\
	&\phantom{=.\frac{1}{2}i}\left\lbrace\frac{1}{8}\left[\eth\left(3h\bar{\eth}\dot{\bar{h}}-3\dot{h}\bar{\eth}\bar{h}+\dot{\bar{h}}\bar{\eth}h-\bar{h}\bar{\eth}\dot{h}\right)\right]\right\rbrace,
	\end{align}
\end{subequations}
and
\begin{subequations}
	\begin{align}
	\label{eq:m}
	m&=-\text{Re}\left[\Psi_{2}+\frac{1}{4}\dot{h}\bar{h}\right],\\
	\label{eq:N}
	\widehat{N}&=2\Psi_{1}-\frac{1}{4}\bar{h}\eth h-u\eth m-\frac{1}{8}\eth(h\bar{h}).
	\end{align}
\end{subequations}
The reason why we construct these contributions to $J$ is because of their parity and type of memory contribution, which we list in Table~\ref{tab:Jtypes}. Note that one can eliminate the $\Psi_{1}$ term in Eq.~\eqref{eq:N} by using one of the Bianchi identities for the Weyl scalars, which produces the relation
\begin{align}
  \label{eq:EliminatePsi1}
\dot{\Psi}_{1}=-\frac{1}{2}\eth\Psi_{2}+\frac{1}{4}\bar{h}\eth\dot{h}.
\end{align}
\begin{table}
	\label{tab:Jtypes}
	\centering
	\caption{The four contributions to $J$ in terms of their parity, type of memory contribution, and interpretation in terms of more common quantities in general relativity.
	}
	\begin{tabular}{cccc}
		\Xhline{3\arrayrulewidth}
		Variable & Parity & Memory & Interpretation\\
		\colrule
		$J_{m}$ & Electric & Ordinary & Bondi mass aspect \\
		$J_{\mathcal{E}}$ & Electric & Null & Energy flux \\
		$J_{\widehat{N}}$ & Magnetic & Ordinary & Angular momentum aspect \\
		$J_{\mathcal{J}}$ & Magnetic & Null & Angular momentum flux \\
		\Xhline{3\arrayrulewidth}
	\end{tabular}
\end{table}

We now use these observables to examine whether the extrapolated strain waveforms actually capture all terms in Eq.~\eqref{eq:J} except $J_{\mathcal{E}}$. We do this by adding $J_{\mathcal{E}}$ and checking the violation of the BMS balance laws, Eq.~\eqref{eq:hJ}, and comparing to CCE waveforms that already capture the gravitational memory effects.

%%%%%%%%%%%%%%%%%%%%%%%%%%%%%%%%%%%%%%%%%%%%%%%%%%%%%%%%%%%
\section{Results}
%%%%%%%%%%%%%%%%%%%%%%%%%%%%%%%%%%%%%%%%%%%%%%%%%%%%%%%%%%%
\label{sec:results}
For the following results, we numerically evolved a set of 13 binary black hole (BBH) mergers with various mass ratios and spin configurations using the Spectral Einstein Code (SpEC)~\cite{SpECCode}. We list the parameters of these evolved BBH systems in Table~\ref{tab:runs}. Each BBH simulation contains roughly 19 orbits prior to merger and is evolved until the gravitational waves from ringdown leave the domain. Unlike evolutions in the SXS catalog, the full set of Weyl scalars and the strain have been extracted from these runs and the asymptotic waveforms have been computed using both the extrapolation technique described in~\cite{Iozzo2020} and the CCE procedure described in~\cite{moxon2020improved,moxon2021}. Extrapolation is performed using the python module \texttt{scri}~\cite{scri_url, Boyle2013, Boyle2016, Boyle2014} and CCE is performed using SpECTRE's CCE code~\cite{moxon2020improved,moxon2021,CodeSpECTRE}. For the CCE extractions, the four world tube radii are chosen to be equally spaced between $2\lambdabar_{0}$ and $21\lambdabar_{0}$, where $\lambdabar_0=1/\omega_0$ is the initial reduced gravitational wavelength as determined by the orbital frequency of the binary from the initial data. These 13 waveforms will be made publicly available in the SXS catalog~\cite{SXSCatalog}.

As mentioned above, our asymptotic strain waveforms are computed using two methods: extrapolation and CCE. The first method uses Regge-Wheeler-Zerilli (RWZ) extraction to compute the strain on a series of concentric spheres of constant coordinate radius and then extrapolates these values to future null infinity using $1/r$ approximations~\cite{Sarbach2001,Regge1957,Zerilli1970,Boyle_2019,Iozzo2020, Boyle_2009}.  This is the strain that can be found in the SXS catalog. The other and more faithful extraction method, known as CCE, computes the strain by using the world tube data provided by a Cauchy evolution as the inner boundary data for a nonlinear evolution of the Einstein field equations on null hypersurfaces~\cite{moxon2020improved,moxon2021}. CCE requires freely specifying the strain on the initial null hypersurface labeled $u=0$. Like~\cite{mitman2020computation}, we choose this field to match the value and the first radial derivative of $h$ from the Cauchy data on the world tube, using the ansatz,
\begin{align}
h(u=0,r,\theta^{A})=\frac{A(\theta^{A})}{r}+\frac{B(\theta^{A})}{r^{3}},
\end{align}
where the two coefficients $A(\theta^{A})$ and $B(\theta^{A})$ are fixed by the Cauchy data on the world tube. Note that constructing a satisfactory initial null hypersurface for CCE is currently an open problem in numerical relativity. Consequences of this choice manifest as transient effects appearing at early times~\cite{mitman2020computation}. This hypersurface choice also determines the Bondi frame of the resulting asymptotic waveform. 

As a result of these junklike transient effects in the CCE waveforms, we cannot expect our extrapolated and CCE waveforms to be in the exact same Bondi frame. Two strain waveforms from the same physical system but in different Bondi frames have infinite degrees of freedom relating them. Therefore, it is only meaningful to compare waveforms if they are in the same Bondi frame. Fortunately, we can numerically apply a BMS transformation to our waveforms until they are in approximately the same Bondi frame~\cite{Boyle2016}. Thus, for any comparisons between extrapolated and CCE waveforms we use an optimization to find the supertranslation that minimizes the $L^{2}$ norm\footnote{Recall that the $L^{2}$ norm of a function $h(\Omega)$ is given by
\begin{align}
||h(\Omega)||\equiv\int|h(\Omega)|^{2}d\Omega.
\end{align}} of the difference of the two strains. For this optimization, the $\ell\leq2$ modes of the supertranslation are free parameters, while the $\ell>2$ modes are set to zero since their inclusion tends to produce negligible changes.\footnote{This is because most of the transient effects that are present in CCE waveforms, which are the main causes of the misalignment, primarily manifest in the $\ell=2$ modes.}

Last, we note that both our extrapolated and CCE waveforms have been post-processed so that they are approximately in the center-of-mass frame~\cite{PhysRevD.100.124010}.

%%%%%%%%%%%%%%%%%%%%%%%%%%%%%%%%%%%%%%%%%%%%%%%%%%%%%%%%%%%
\subsection{Issues with Extrapolated Waveforms}
%%%%%%%%%%%%%%%%%%%%%%%%%%%%%%%%%%%%%%%%%%%%%%%%%%%%%%%%%%%
\label{sec:extrap}
\begin{figure}
	\label{fig:ExtBMS_2_0}
	\centering
	\includegraphics[width=\columnwidth]{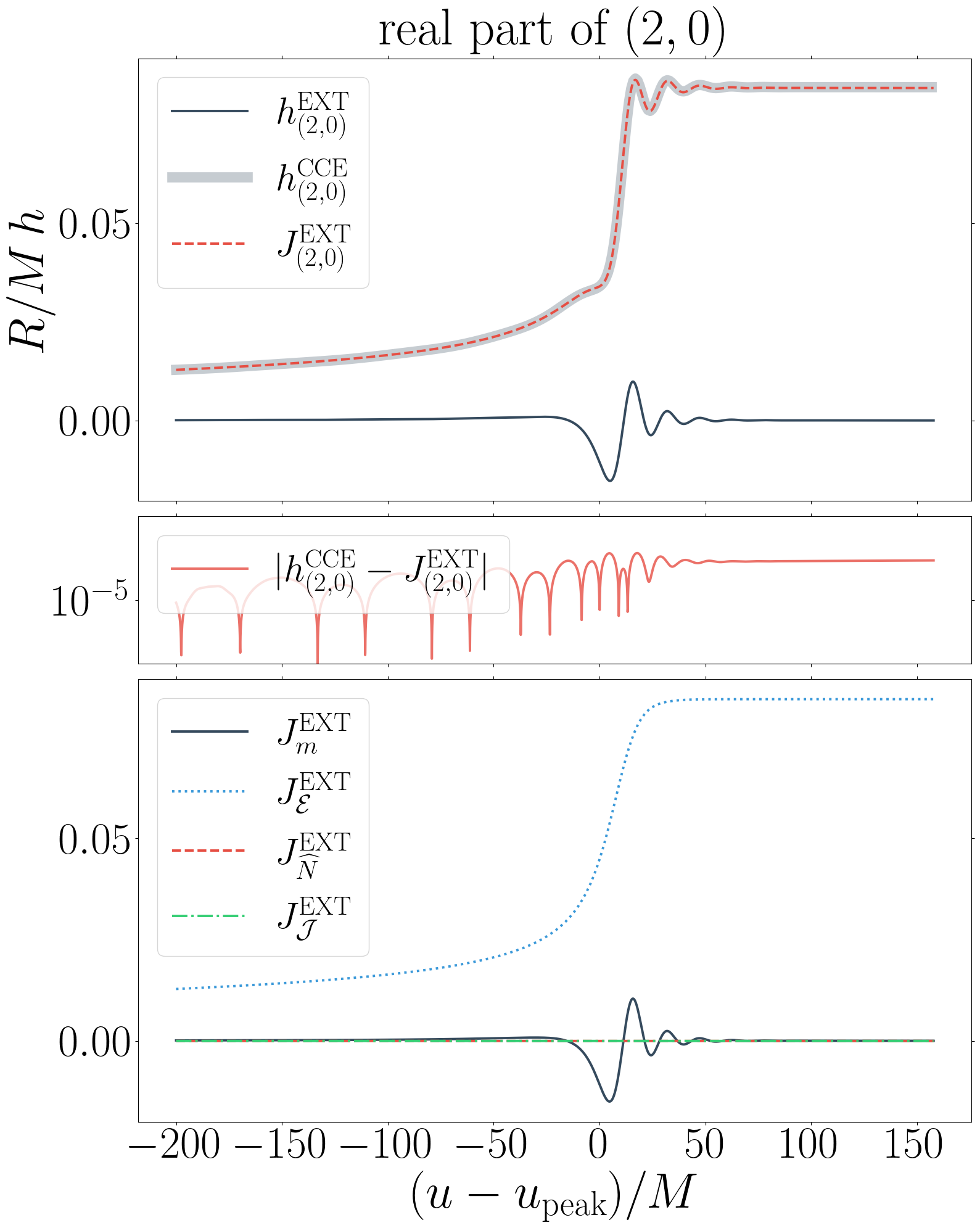}
	\caption{Comparison of the extrapolated strain $(2,0)$ mode ($h^{\text{EXT}}$, black/solid) to the strain that is extracted using CCE ($h^{\text{CCE}}$, gray/solid) and to what is expected according to the other modes of the extrapolated waveform using the various BMS balance laws ($J^{\text{EXT}}$, red/dashed) computed by using the extrapolated waveform in Eq.~\eqref{eq:Js}. We take the news to be the retarded time derivative of the strain. In the middle panel, we plot the absolute error: $|h^{\text{CCE}}-J^{\text{EXT}}|$. In the bottom panel, we outline the individual contributions that come from the Bondi mass aspect (black/solid), the Bondi angular momentum aspect (red/dashed), the energy flux (blue/dotted), and the angular momentum flux (green/dashed/dotted).\\
	BBH merger: \texttt{q1\_nospin} (see Table~\ref{tab:runs}).\\
	CCE waveform: CCE-R0292 (with a supertranslation applied).}
\end{figure}

We first illustrate the extrapolated waveforms' inability to capture the energy flux term $J_{\mathcal{E}}$ in comparison to CCE and the strain from the BMS balance laws. As shown in the top panel of Fig.~\ref{fig:ExtBMS_2_0}, the extrapolated strain waveform's $(2,0)$ mode is constant except for quasinormal mode oscillations near merger and ringdown. However, the strain that is computed from the BMS balance laws $J$ not only contains these quasinormal mode oscillations, but also a contribution from the growth of the memory. Further, the bottom panel of Fig.~\ref{fig:ExtBMS_2_0} shows that the displacement memory contribution to $J$ predominantly comes from the energy flux term $J_{\mathcal{E}}$. The dominance of $J_{\mathcal{E}}$ should not come as a surprise since it has been shown that the ordinary contributions to the displacement memory, $J_{m}$ and $J_{\widehat{N}}$, should be negligible for BBH mergers~\cite{ashtekar2019compact,PhysRevD.101.044005}, and the contribution from $J_{\mathcal{J}}$ should be zero at late times since the news vanishes for Kerr spacetimes. What is perhaps surprising, though, is that the BMS balance law strain $J$ that we compute from the extrapolated waveform is nearly identical to the CCE strain. We explore this comparison between extrapolated and CCE waveforms more thoroughly in Sec.~\ref{sec:ccecomparison}. Note that for the plot in Fig.~\ref{fig:ExtBMS_2_0} and all of the following plots we take $u_{\text{peak}}$ to be the time at which the $L^{2}$ norm of the extrapolated strain achieves its maximum value.

\begin{figure}
	\label{fig:BMS_Violation}
	\centering
	\includegraphics[width=\columnwidth]{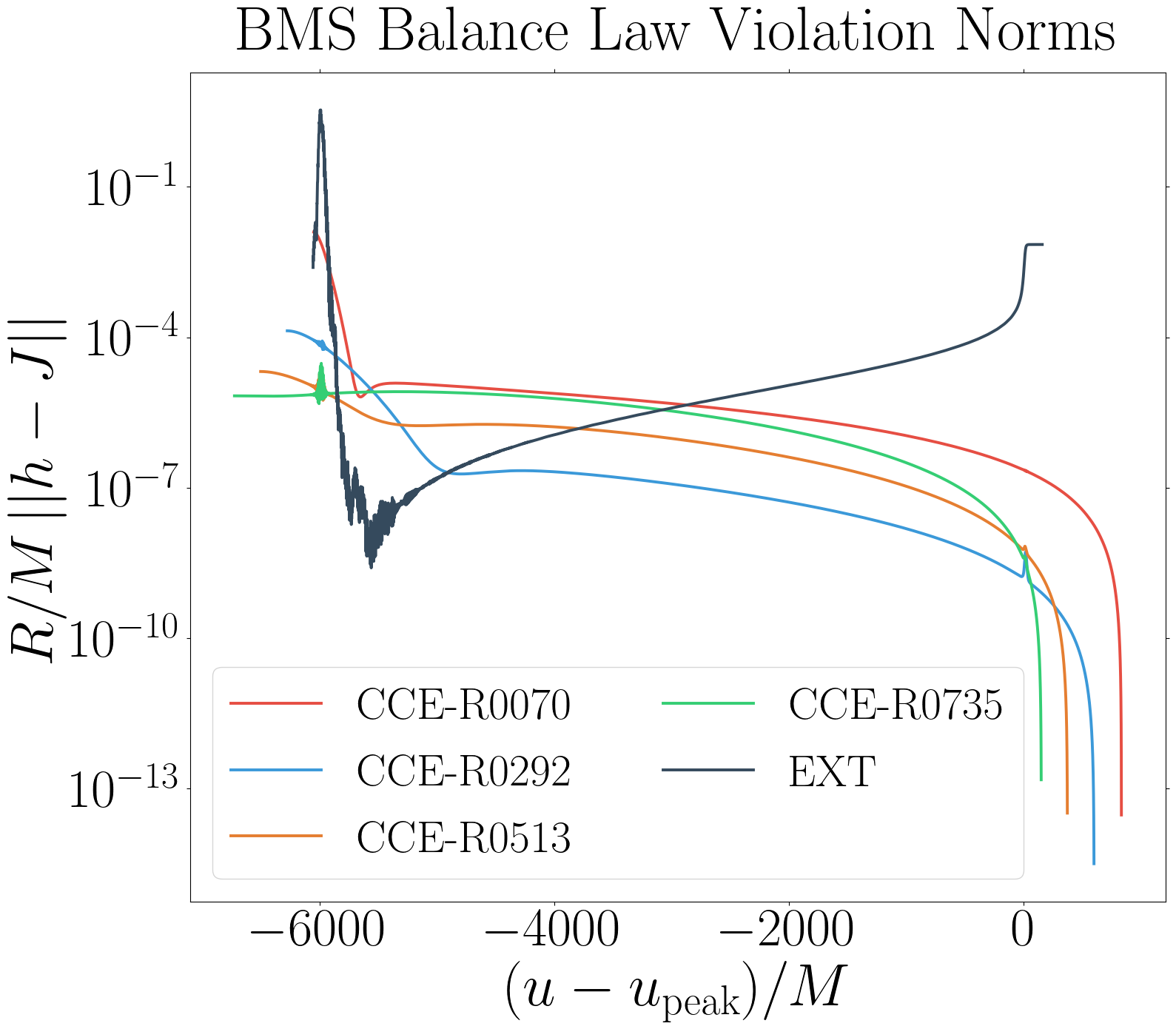}
	\vspace*{0.25cm}
	\includegraphics[width=\columnwidth]{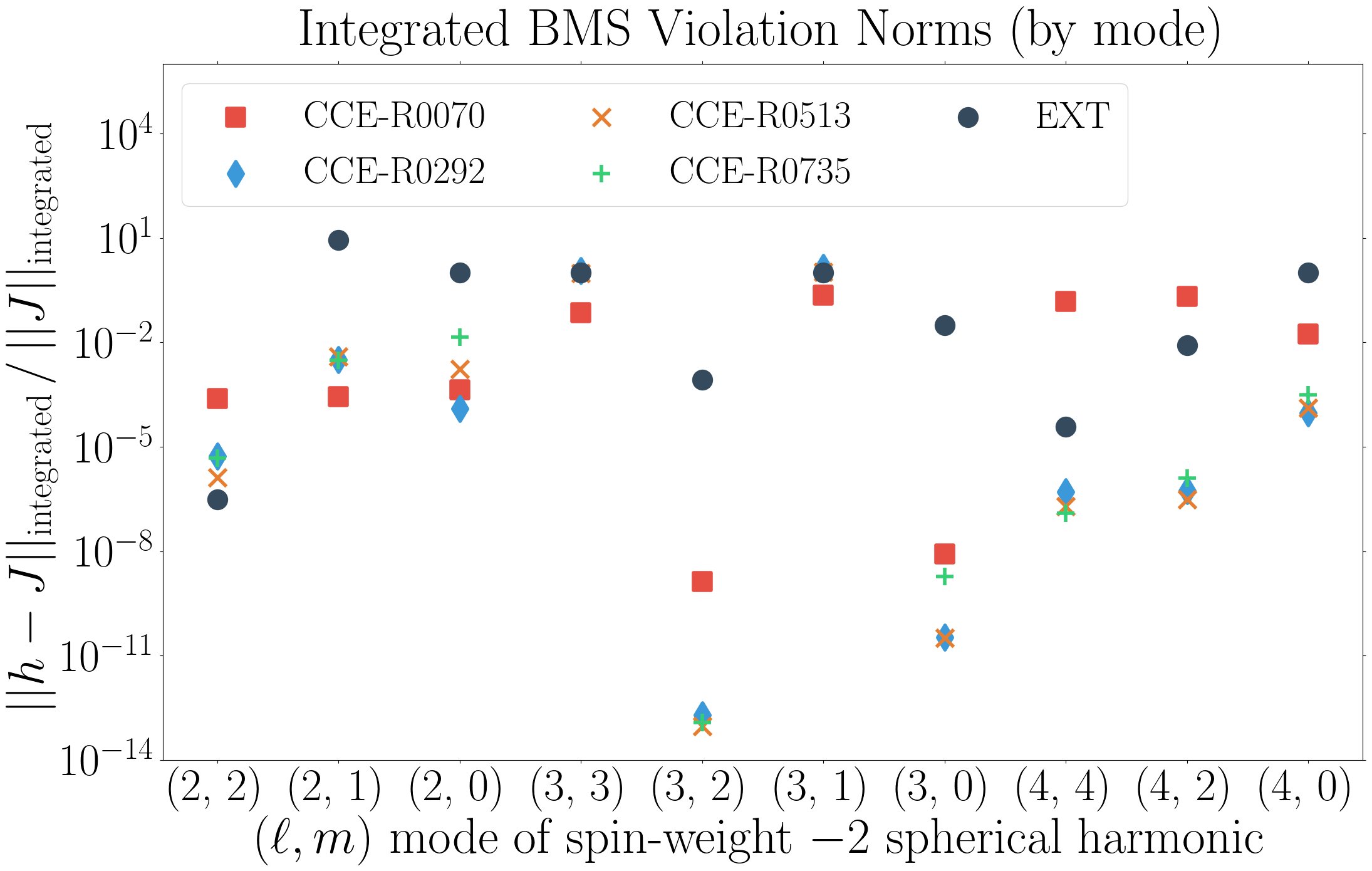}
	\caption{Violation of the BMS balance laws, Eq.~\eqref{eq:hJ}, for extrapolated and CCE waveforms. In the top plot we show the norm of Eq.~\eqref{eq:hJ} over the two-sphere. Here $J$ in Eq.~\eqref{eq:hJ} is computed by Eq.~\eqref{eq:J} and the $\Psi_2$ used in Eq.~\eqref{eq:J} for the extrapolated or CCE waveforms is extracted using the procedures described in~\cite{Iozzo2020} and~\cite{moxon2020improved,moxon2021}. In the bottom plot we show the normalized time-integrated $L^{2}$ norm as a function of a few important spin-weight $-2$ spherical harmonic modes. Note that we integrate each of these modes by starting at one full orbit past the retarded time $u=0\,M$ to suppress misleading effects that the Cauchy evolution junk radiation or the CCE transient effects may induce.\\
	BBH merger: \texttt{q1\_nospin} (see Table~\ref{tab:runs}).
	}
\end{figure}

Because of the extrapolated waveforms' inability to capture the energy flux contribution, we expect their violation of the BMS balance laws, Eq.~\eqref{eq:hJ}, to be much more significant than for waveforms extracted using CCE. In Fig.~\ref{fig:BMS_Violation} we plot the violation of the BMS balance laws over the two-sphere (top plot) and as a function of certain modes (bottom plot) for the extrapolated waveforms and the CCE waveforms corresponding to the four available world tube extraction radii. The values that we plot in the bottom plot are the normalized time integrals of the BMS balance law violation from one full orbit past the retarded time $u=0\,M$ onward. We exclude the first orbit to suppress misleading effects that the Cauchy evolution junk radiation or the CCE transient effects may induce~\cite{Boyle_2019,mitman2020computation}. As can be seen in the top plot, the violation of the BMS balance laws by the extrapolated waveforms is roughly two orders of magnitude more than the worst-performing CCE waveform (R0070) and four orders of magnitude more than the best-performing CCE waveform (R0292). Moreover, in the bottom plot of Fig.~\ref{fig:BMS_Violation}, one can easily observe that sources of this violation are predominantly from the $m=0$ primary memory modes, as is perhaps expected.

\subsection{Correcting Extrapolated Waveforms}
%%%%%%%%%%%%%%%%%%%%%%%%%%%%%%%%%%%%%%%%%%%%%%%%%%%%%%%%%%%
\label{sec:extrapwithmemory}

Having thoroughly described the problems with the extrapolated waveforms, we now discuss our method for ``adding memory'' to these waveforms. We then reevaluate their violation of the BMS balance laws and compare them to the CCE waveforms.

\subsubsection{Adding Memory to Extrapolated Waveforms}

As has been shown analytically in~\cite{ashtekar2019compact,PhysRevD.101.044005} and also numerically in~\cite{mitman2020computation}, the ordinary and magnetic contributions to the displacement memory should be negligible, if not vanish completely. In other words, the net change in the function $J$ should be almost entirely sourced by $J_{\mathcal{E}}$, the energy flux contribution. In agreement with our hypothesis, Fig.~\ref{fig:ExtBMS_2_0} shows that the extrapolated waveform is indeed reasonably modeled by just the $J_{\Psi}$ term, thereby implying that $J_{\mathcal{E}}$ provides the missing time evolution of the extrapolated strain. Consequently, since $J_{\mathcal{E}}$ is a function only of the news, we can recover the full time dependence of the displacement memory's growth.

Following the works of~\cite{mitman2020computation, Talbot_2018}, we now use Eq.~\eqref{eq:JE} to compute the time evolution of the displacement memory in the extrapolated strain. To avoid any memory effects that are induced by junk radiation, we do not time integrate the entire waveform. We instead take the lower limit $u_{1}$ to be half of one orbit past the retarded time $u=0\,M$, which roughly matches up with the relaxation time, i.e., the time at which the junk radiation from the Cauchy evolution is considered to be negligible~\cite{Boyle_2019}. Unlike~\cite{Talbot_2018}, after computing $J_{\mathcal{E}}$, we then add this contribution back to the strain to produce a more BMS-accurate waveform.

\begin{figure}
	\label{fig:BMS_Violation_withmem}
	\centering
	\includegraphics[width=\columnwidth]{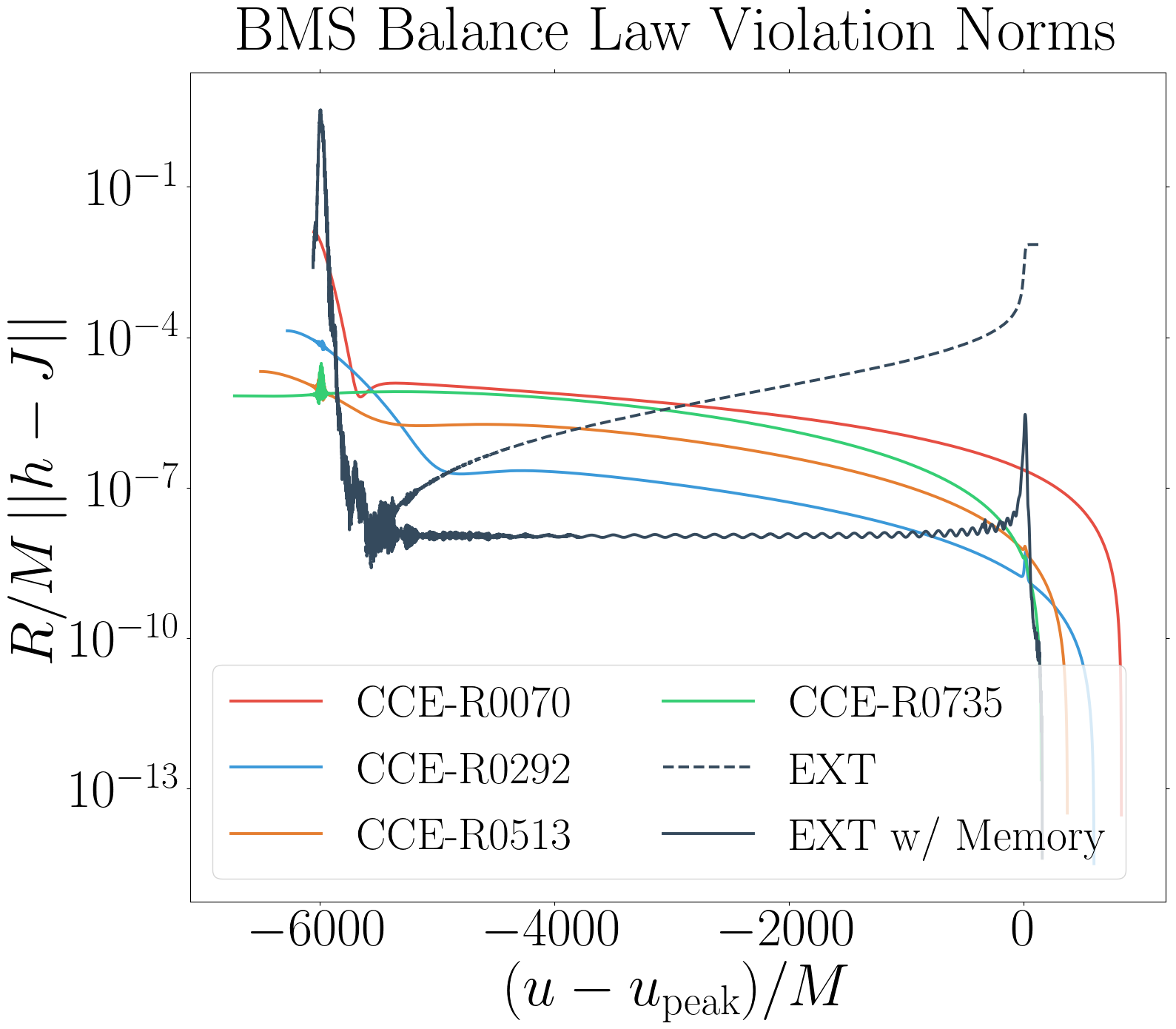}
	\vspace*{0.25cm}
	\includegraphics[width=\columnwidth]{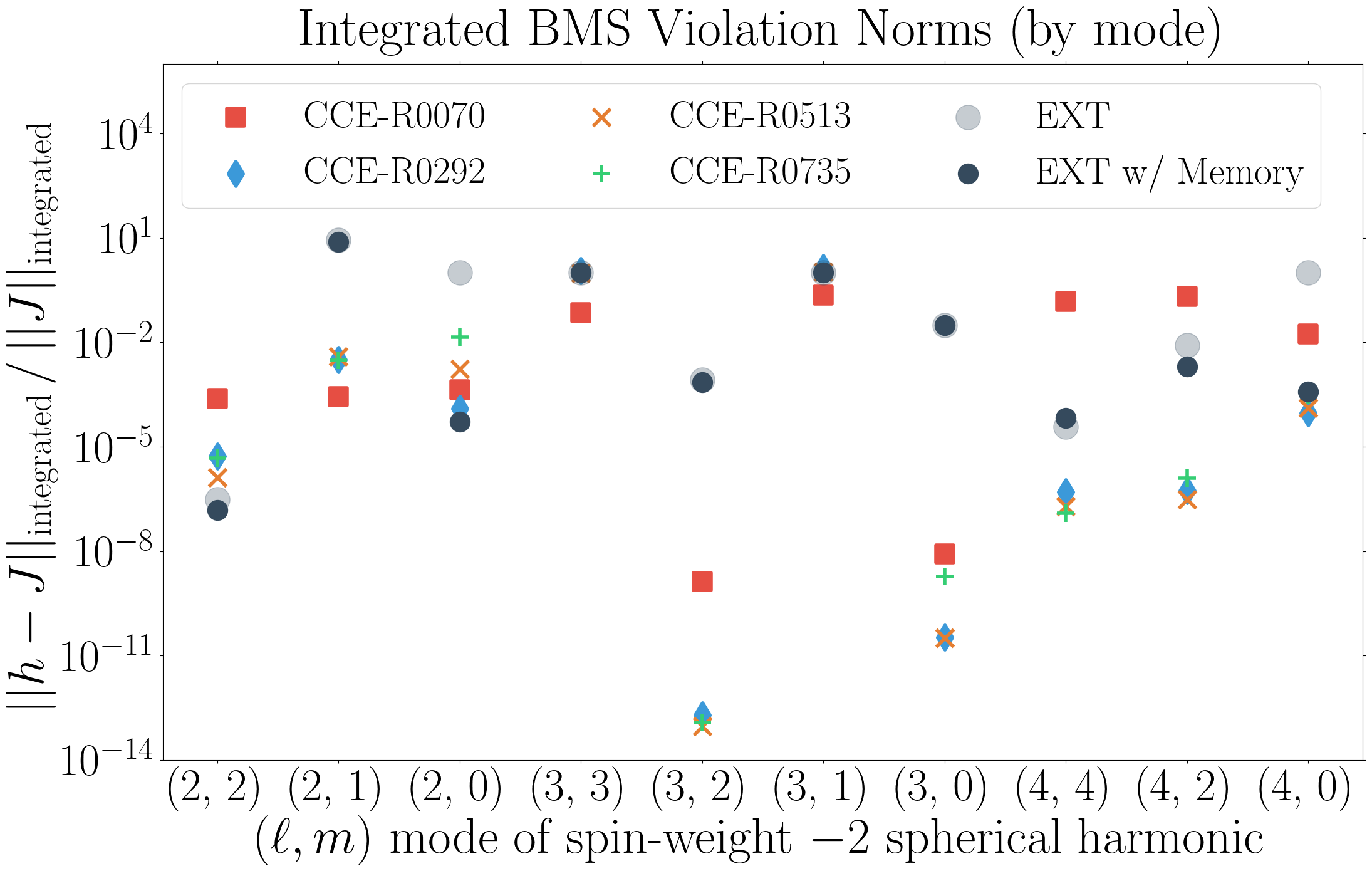}
	\caption{Identical to Fig.~\ref{fig:BMS_Violation}, but now with the violations for the memory-corrected extrapolated strain.\\
	BBH merger: \texttt{q1\_nospin} (see Table~\ref{tab:runs}).}
\end{figure}
\begin{figure}
	\label{fig:BMS_Violation_withmem_precessing}
	\centering
	\includegraphics[width=\columnwidth]{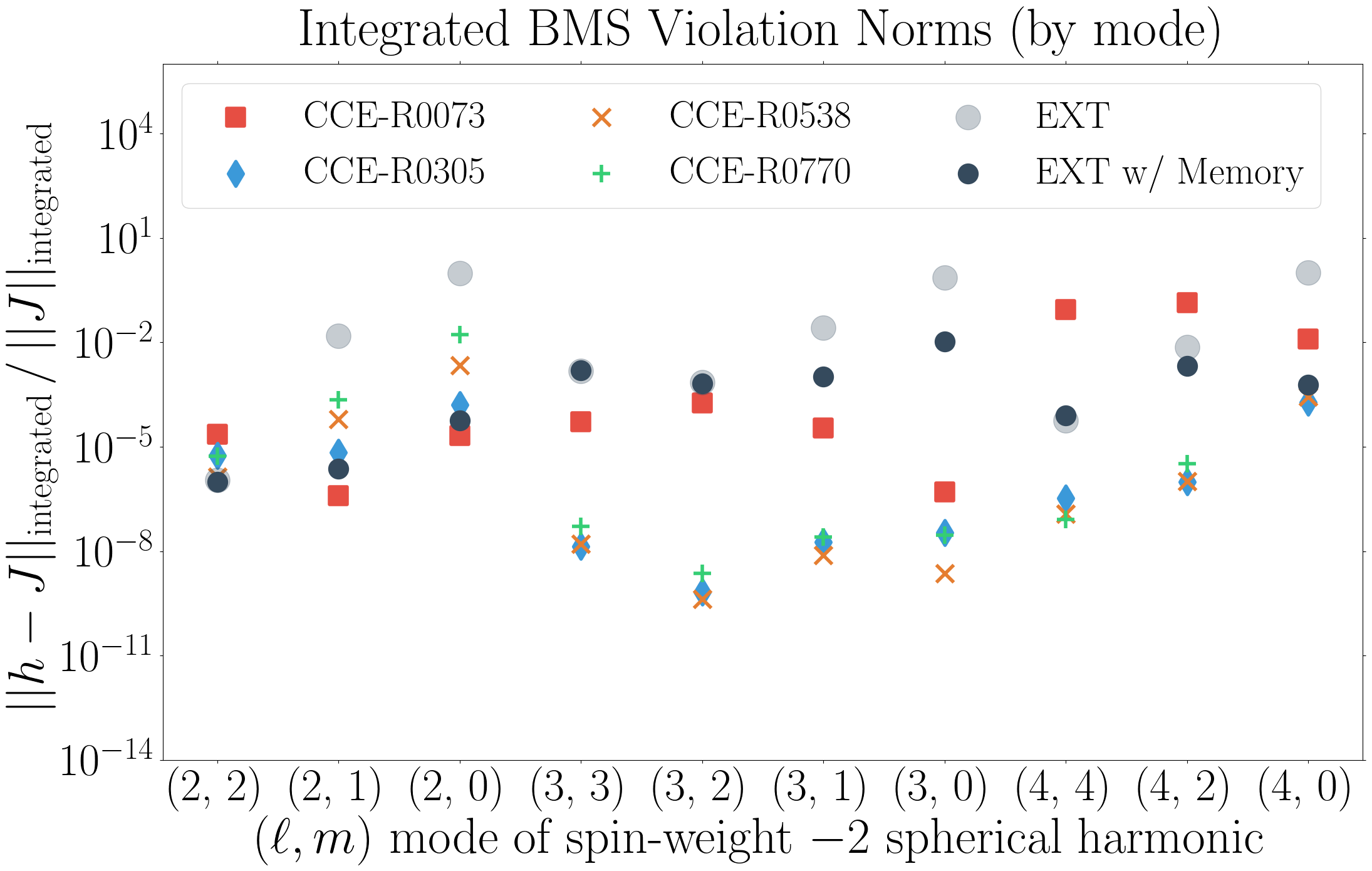}
	\caption{Identical to the bottom plot of Fig.~\ref{fig:BMS_Violation_withmem}, but now for an equal mass, precessing system.\\
	BBH merger: \texttt{q1\_precessing} (see Table~\ref{tab:runs}).}
\end{figure}

In Fig.~\ref{fig:BMS_Violation_withmem} we show the violation of the BMS balance laws by the memory-corrected extrapolated strain over the two-sphere and as a function of certain modes. Again, the values that we plot in the bottom plot are the normalized time integrals of this violation from one full orbit past the retarded time $u=0\,M$ onward. As expected, the overall violation of the BMS balance laws over the two-sphere improves by nearly four orders of magnitude, with the major improvements seen in the $\ell=\text{even},m=0$ memory modes, but also in unexpected modes, such as the $(2,2)$ and $(4,2)$ modes. For a precessing system, the improvement can be seen in many more modes, as shown in Fig.~\ref{fig:BMS_Violation_withmem_precessing}. We choose to show the results for the equal mass precessing system, rather than the $q=4$ precessing system, because the memory is known to increase as the mass ratio approaches unity~\cite{mitman2020computation, Pollney_2011}.  We note that for evaluating the balance laws it is important to compute the BMS strain $J$ from Eq.~\eqref{eq:J} rather than Eq.~\eqref{eq:Js} since the Weyl scalar $\Psi_{1}$ tends to make extrapolated waveform computations of $J$ noticeably worse, in terms of the balance law violation, because of spurious effects induced by junk radiation.
\begin{figure*}
	\label{fig:BMS_Violation_All}
	\centering
	\includegraphics[width=\textwidth]{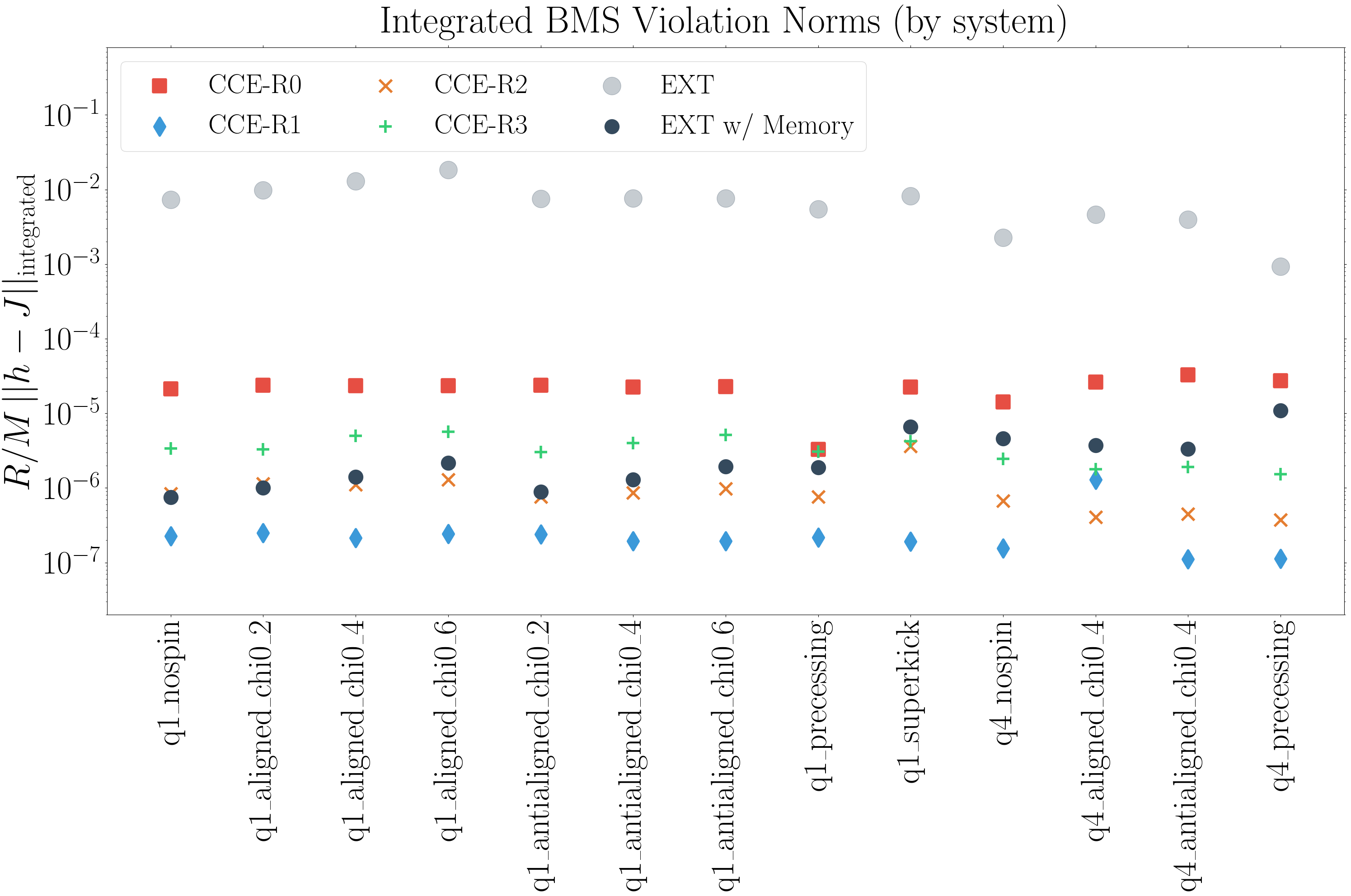}
	\caption{The time-integrated norm of the BMS balance law violation as a function of the BBH system. Integration is performed from one full orbit past the retarded time $u=0\,M$ onward to avoid errors introduced by the Cauchy evolution junk radiation and most of the CCE transient effects. The labeling CCE-R[0,1,2,3] represents the four world tube radii (from smallest to largest). This plot demonstrates that the displacement memory can be effectively added to the extrapolated waveforms with the result satisfying the BMS balance laws to comparable accuracy as CCE. The parameters of these systems can be found in Table~\ref{tab:runs}.}
\end{figure*}

Figure~\ref{fig:BMS_Violation_All} shows the time integral of the norm of the BMS balance law over the two-sphere for all the systems listed in Table~\ref{tab:runs}, both before and after the memory corrections to the extrapolated strain waveforms. As can be seen, for every type of BBH merger that we present, the improvements to the extrapolated waveforms are quite considerable.

A few of the extrapolated waveforms in Fig.~\ref{fig:BMS_Violation_All} seem to be better than the corresponding CCE waveforms. This is simply because our time-integration range is chosen to ignore the Cauchy evolution's junk radiation completely, but not all of the longer-lasting transients caused by imperfect CCE initial null hypersurface data. If we instead choose $u_1$ to be later in the inspiral, then these CCE waveforms also outperform the corrected extrapolated waveforms. One can easily see this by comparing the violations for extrapolation and CCE around the time of merger in, for example, the top plot of Fig.~\ref{fig:BMS_Violation_withmem}.

\subsubsection{Limitations of the Corrected Waveforms}

As one may have noticed in the top plot of Fig.~\ref{fig:BMS_Violation_withmem}, while the overall violation of the BMS balance laws is improved by including the energy flux contribution, near merger the extrapolated waveforms' violation is much larger than that of the various CCE waveforms. The reason for this is that, while we have corrected the extrapolated waveforms by adding just $J_{\mathcal{E}}$, we have done nothing regarding the magnetic terms $J_{\widehat{N}}$ and $J_{\mathcal{J}}$. These magnetic terms are, in part, the source of the spin memory effect when the strain is integrated with respect to retarded time~\cite{Pasterski_2016,Nichols_2017,Comp_re_2020}. The relevance of the magnetic component can be seen by noticing that the $(3,0)$, $(2,1)$, and $(3,2)$ modes---which are a few of the primary contributors to the spin memory---are the largest sources of the BMS balance law violation, even after the displacement memory correction is applied. (See, e.g., the bottom plot of Fig.~\ref{fig:BMS_Violation_withmem}.)

\begin{figure}
	\label{fig:ExtBMS_3_0_imag}
	\centering
	\includegraphics[width=\columnwidth]{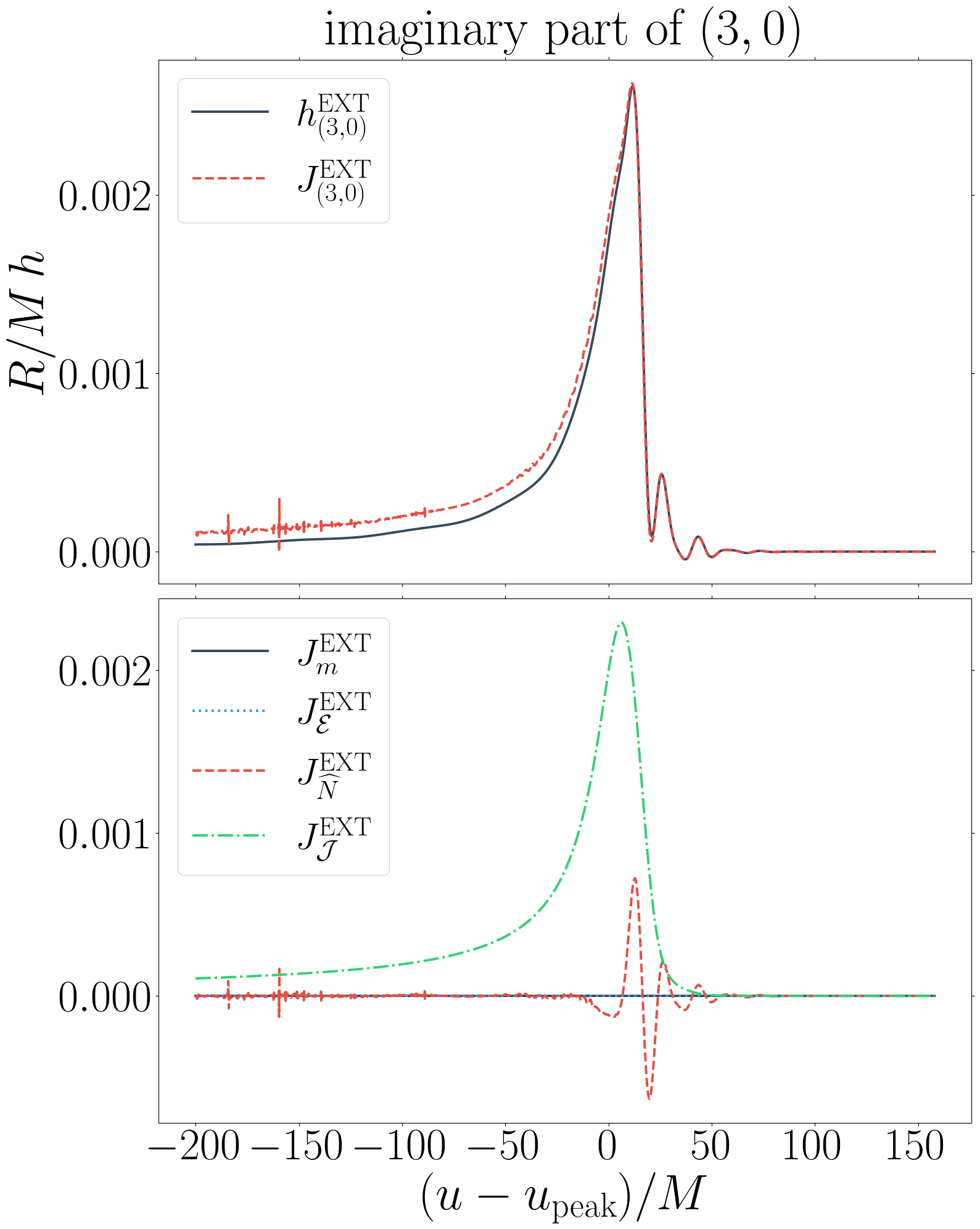}
	\caption{Comparison of the extrapolated strain $(3,0)$ mode ($h^{\text{EXT}}$, black/solid) to the BMS strain that we compute using the terms in Eq.~\eqref{eq:Js} ($J^{\text{EXT}}$, red/dashed). In the bottom panel, we display the contributions that come from the Bondi mass aspect (black/solid), the Bondi angular momentum aspect (red/dashed), the energy flux (blue/dotted), and the angular momentum flux (green/dashed/dotted).\\
		BBH merger: \texttt{q1\_nospin} (see Table~\ref{tab:runs}).}
\end{figure}

As has been shown previously~\cite{mitman2020computation}, even though the extrapolated waveforms capture the spin memory effect---which we recall is a memory effect in the time integral of the strain---for some unknown reason the magnitude of the spin memory is roughly 50\% of what it should be according to comparisons with the BMS balance laws and CCE waveforms. In Fig.~\ref{fig:ExtBMS_3_0_imag}, we compare the strain of the extrapolated waveform to the strain obtained from the BMS balance laws for the main spin memory mode: the imaginary part of the $(3,0)$ mode. As can readily be seen in the lower panel, the BMS strain is primarily sourced by the $J_{\widehat{N}}$ and $J_{\mathcal{J}}$ terms, with the contribution of $J_{\widehat{N}}$ representing the oscillatory part and the contribution of $J_{\mathcal{J}}$ representing the time derivative of the nonoscillatory spin memory. However, unlike the displacement memory mode, cf. Fig.~\ref{fig:ExtBMS_2_0}, which is not present in the extrapolated waveforms, the top panel of Fig.~\ref{fig:ExtBMS_3_0_imag} shows that there is a non-zero contribution to the time derivative of the spin memory. Consequently, without using the Weyl scalar $\Psi_{2}$, there is unfortunately no method to accurately correct the spin memory component of the extrapolated waveforms. While we could make this correction to our limited set of BBH simulations, we could not apply such a correction to the entire SXS catalog, since $\Psi_{2}$ has not been extracted for those simulations.

To provide a measurement of this subtle discrepancy, we compute the mismatch between the extrapolated strain with the memory correction and the BMS strain computed from the extrapolated waveform via
\begin{align}
\label{eq:mismatch}
\mathcal{M}(u,\,&h_{\ell m}^{\text{EXT}}, J_{\ell m}^{\text{EXT}})\equiv\nonumber\\
&1-\frac{\langle h_{\ell m}^{\text{EXT}},J_{\ell m}^{\text{EXT}}\rangle}{\sqrt{\langle h_{\ell m}^{\text{EXT}},h_{\ell m}^{\text{EXT}}\rangle\langle J_{\ell m}^{\text{EXT}},J_{\ell m}^{\text{EXT}}\rangle}},
\end{align}
where the inner product is given by
\begin{align}
\langle h_{\ell m}^{\text{EXT}},J_{\ell m}^{\text{EXT}}\rangle\equiv\int_{u_{1}}^{u} h_{\ell m}^{\text{EXT}}\overline{J_{\ell m}^{\text{EXT}}}\,du.
\end{align}
For an equal mass, non-spinning system the mismatches for the $(2,2)$, $(2,0)$, and $(3,0)$ modes are
\begin{subequations}
\begin{align}
\mathcal{M}(u_{\text{final}}, h_{(2,2)}^{\text{EXT}}, J_{(2,2)}^{\text{EXT}})&=6.84\times10^{-8},\\
\mathcal{M}(u_{\text{final}}, h_{(2,0)}^{\text{EXT}}, J_{(2,0)}^{\text{EXT}})&=2.43\times10^{-5},\\
\mathcal{M}(u_{\text{final}}, h_{(3,0)}^{\text{EXT}}, J_{(3,0)}^{\text{EXT}})&=1.36\times10^{-2}.
\end{align}
\end{subequations}
Thus, we observe that the mismatch for the $(3,0)$ mode is considerably worse than the primary strain mode and the displacement memory mode. Nonetheless, despite the remaining problems with magnetic memory effects, it is clear that the physical accuracy of the extrapolated waveforms can be vastly improved by correcting the strain waveform via the addition of $J_{\mathcal{E}}$, as discussed above.

\subsubsection{Comparisons with CCE}
\label{sec:ccecomparison}

\begin{figure}
	\label{fig:CCE_Comparison}
	\centering
	\includegraphics[width=\columnwidth]{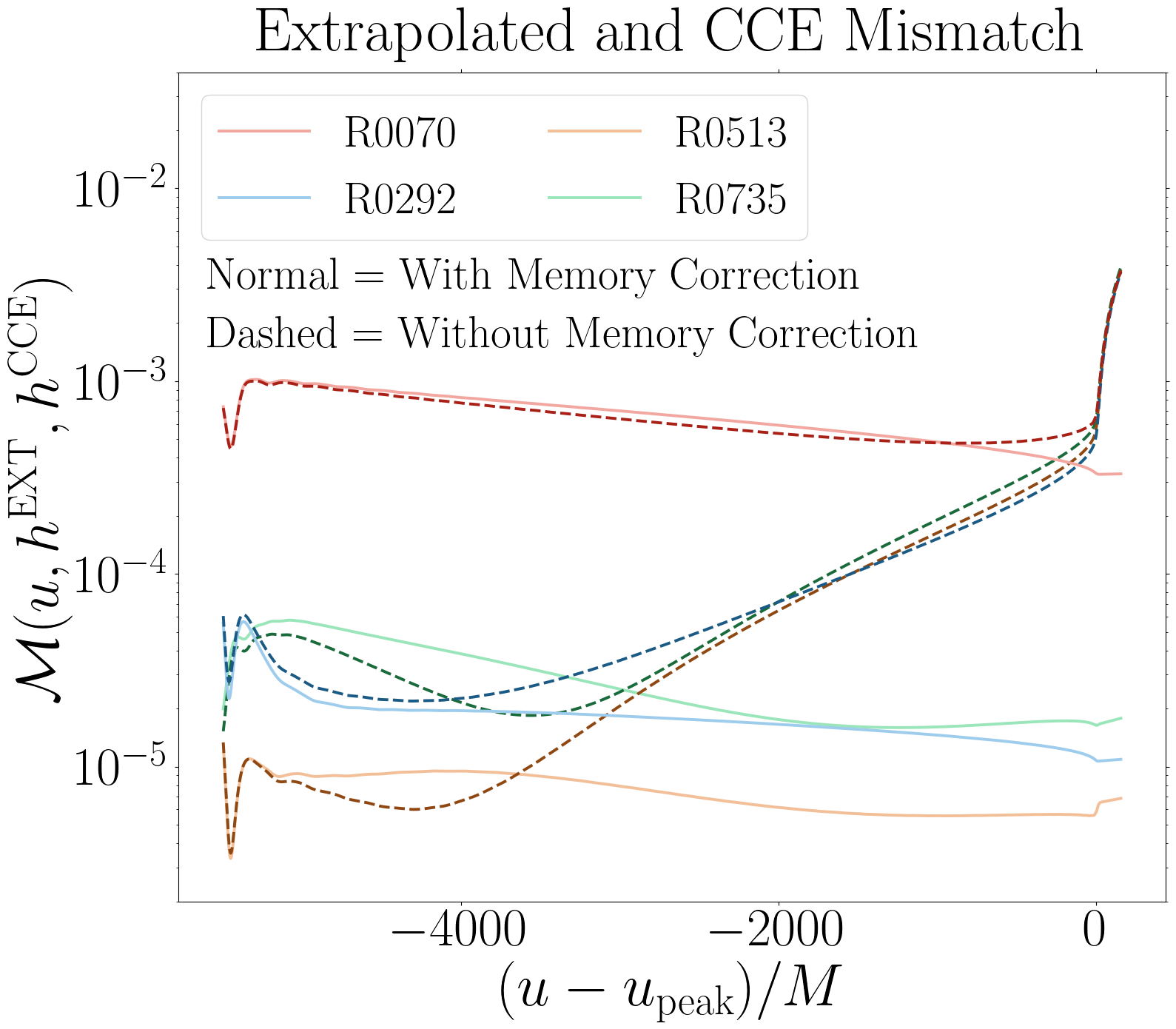}
	\caption{Mismatch between the extrapolated and CCE strains with (light/solid curves) and without (dark/dashed curves) the memory correction applied to the extrapolated strain. We compute the mismatch between the various waveforms according to Eq.~\eqref{eq:mismatchtwosphere}. To improve the alignment between the extrapolated and CCE waveforms, we have applied an optimized $\ell\leq2$ supertranslation to the CCE waveforms.\\
	BBH merger: \texttt{q1\_nospin} (see Table~\ref{tab:runs}).}
\end{figure}

\begin{figure}
	\label{fig:CCE_Comparison_mode}
	\centering
	\includegraphics[width=\columnwidth]{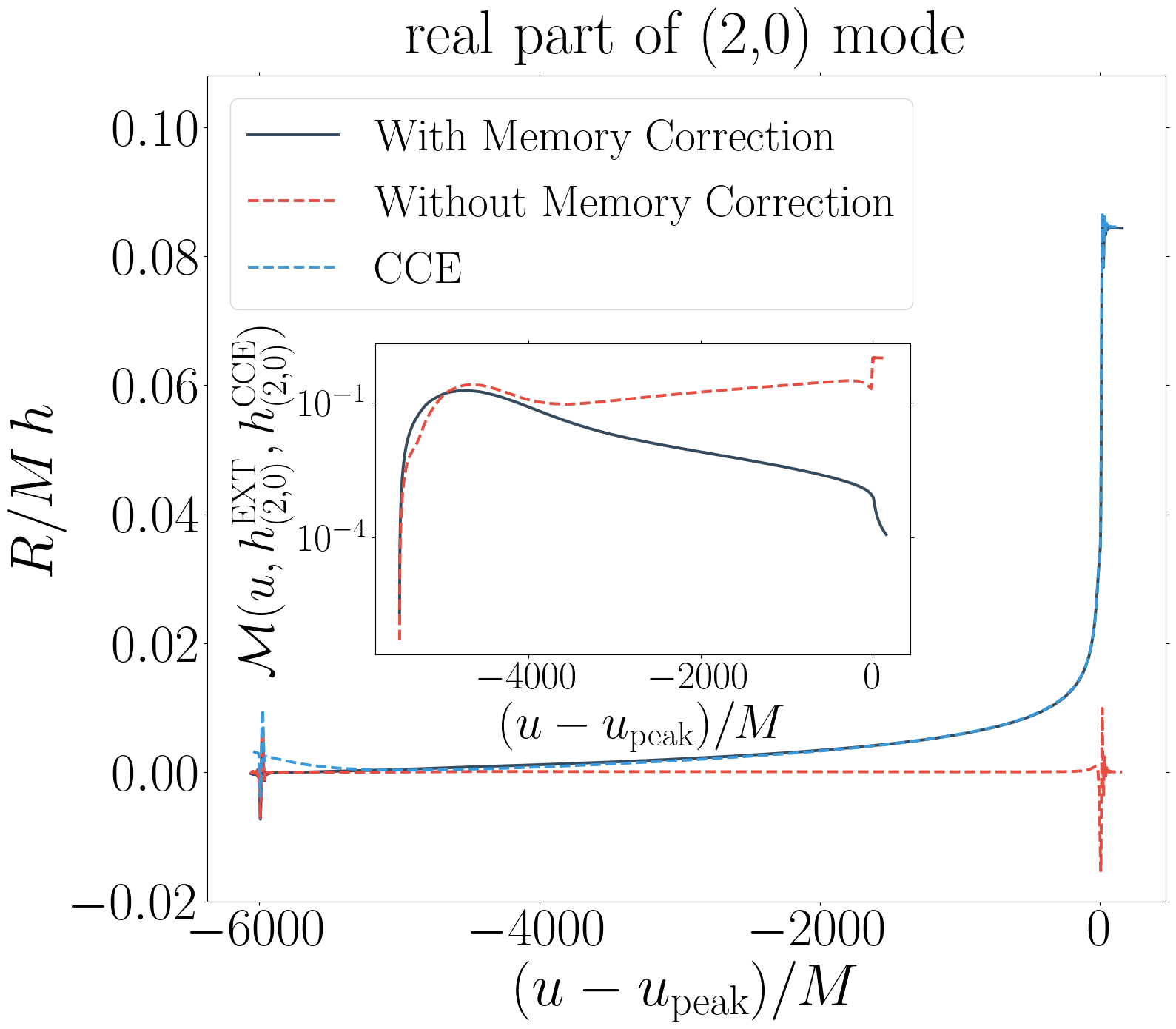}
	\caption{The strain $(2,0)$ mode for the CCE strain waveform (second smallest extraction radius) and the extrapolated strain waveform with and without the memory correction. In the inset panel, we plot the mismatch between the extrapolated and CCE strain waveforms for the same mode using Eq.~\eqref{eq:mismatch}. Further, to improve the alignment between the extrapolated and CCE waveforms, we have applied an $\ell\leq2$ supertranslation to the CCE waveforms that minimizes the $L^{2}$ norm between the extrapolated and CCE waveforms. The negative slope in the CCE strain around $-6000\,M$ is an example of the junklike transient effects in CCE waveforms~\cite{mitman2020computation}.\\
		BBH merger: \texttt{q1\_nospin} (see Table~\ref{tab:runs}).}
\end{figure}

Finally, we compare our memory-corrected extrapolated waveforms to those of CCE. Figure~\ref{fig:CCE_Comparison} illustrates the mismatch between the memory-corrected extrapolated strain waveform and the CCE strain waveforms integrated over the two-sphere. Expressed differently, we plot
\begin{align}
\label{eq:mismatchtwosphere}
\mathcal{M}(u,\,&h^{\text{EXT}},h^{\text{CCE}})\equiv\nonumber\\
&1-\frac{\langle h^{\text{EXT}}, h^{\text{CCE}}\rangle}{\sqrt{\langle h^{\text{EXT}}, h^{\text{EXT}}\rangle\langle h^{\text{CCE}}, h^{\text{CCE}}\rangle}},
\end{align}
where the inner product is given by
\begin{align}
\langle h^{\text{EXT}}, h^{\text{CCE}}\rangle\equiv\int_{u_{1}}^{u}\int_{S^{2}}h^{\text{EXT}}\overline{h^{\text{CCE}}}\,d\Omega\,du.
\end{align}
Figure~\ref{fig:CCE_Comparison_mode} shows the same mismatch but in the $(2,0)$ mode only, which is the primary displacement memory mode. For this figure, we compute the mismatch using Eq.~\eqref{eq:mismatch}, but with $J_{\ell m}^{\text{EXT}}$ replaced by $h_{\ell m}^{\text{CCE}}$ instead. Further, for the mode-by-mode comparison we use the CCE waveform corresponding to the second smallest extraction radius, since this waveform appears to be the most accurate in terms of the BMS balance laws. Recall that we also align the extrapolated and CCE waveforms by applying an $\ell\leq2$ supertranslation to the CCE waveforms that minimizes the $L^{2}$ norm of the difference of the two strain waveforms. As can be seen in Fig.~\ref{fig:CCE_Comparison}, the mismatch between the extrapolated and CCE waveforms is roughly the same for the unchanged and memory-corrected strains during the first $1000\,M$ of the inspiral phase. But beyond this point, the mismatch between the waveforms is considerably better when using the memory-corrected extrapolated waveform. As expected, this improvement is primarily due to the smaller mismatch in the $\ell=\text{even},m=0$ modes, as shown for the $(2,0)$ mode in Fig.~\ref{fig:CCE_Comparison_mode}.

%%%%%%%%%%%%%%%%%%%%%%%%%%%%%%%%%%%%%%%%%%%%%%%%%%%%%%%%%%%
\subsection{Measuring the Bondi Frame}
%%%%%%%%%%%%%%%%%%%%%%%%%%%%%%%%%%%%%%%%%%%%%%%%%%%%%%%%%%%
\label{sec:BondiFrame}

Another important and useful aspect of the BMS group in numerical relativity is determining a BBH system's BMS frame.\footnote{Generally the BMS frame refers to the asymptotic frame that is defined by all BMS freedoms, while the Bondi frame refers to the frame that is defined just by supertranslation freedoms.} The SXS catalog already provides waveforms that have a center-of-mass correction~\cite{PhysRevD.100.124010,Boyle_2019}. Yet, just as space and time translation symmetries correspond to four-momentum, the symmetries of the supertranslations correspond to an infinite-dimensional group of conserved charges that are called the \emph{supermomentum}. Further, analogous to how the four-momentum can be used to define a unique frame, such as the rest frame, up to an arbitrary spacetime translation, so can the supermomentum be used to define a unique Bondi frame up to some spacetime translation. This frame is known in the literature as the \emph{nice section} or the \emph{super rest frame}.

While there are many different constructions of the supermomentum~\cite{Dain2000, Dray1984, Geroch1977, Geroch1981, ashtekar1981symplectic}, the one that corresponds to the Bondi frame of an asymptotic waveform is understood to be the Moreschi supermomentum~\cite{Moreschi_2004},
\begin{align}
\label{eq:Moreschi}
\Psi^{\text{M}}_{\ell m}\equiv-\frac{1}{\sqrt{4\pi}}\int_{S^2}Y_{\ell m}\left[\Psi_{2}+\frac{1}{4}\dot{h}\bar{h}+\frac{1}{4}\eth^2 h\right]d\Omega,
\end{align}
where $S^2$ represents the two-sphere.

\begin{figure}
	\label{fig:supermomentum}
	\centering
	\includegraphics[width=\columnwidth]{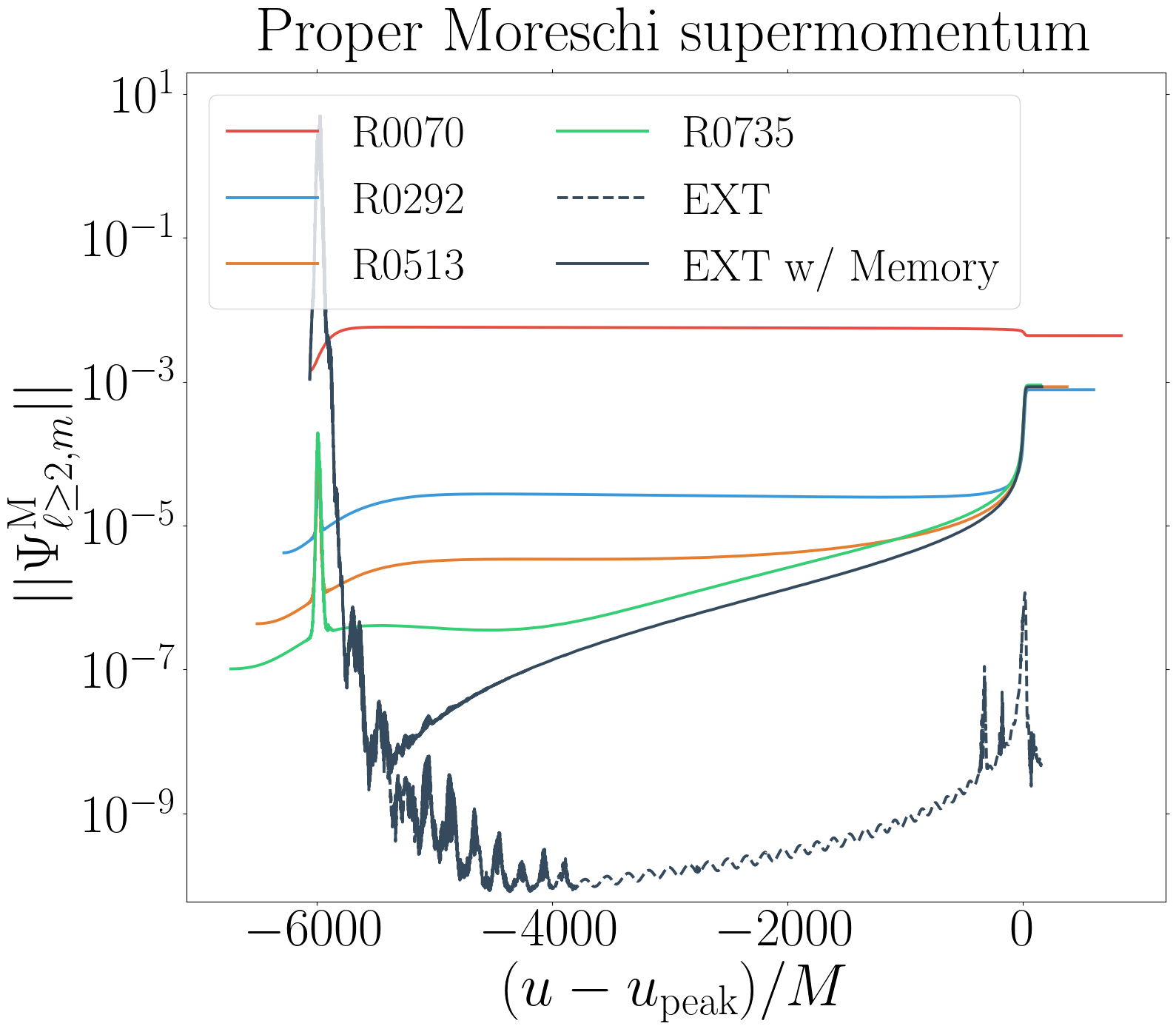}
	\caption{Norm of the proper Moreschi supermomentum, as defined by Eq.~\eqref{eq:Moreschi}, for the extrapolated waveforms, both with and without memory corrections, and the CCE waveforms.\\
		BBH merger: \texttt{q1\_nospin} (see Table~\ref{tab:runs}).}
\end{figure}

In Fig.~\ref{fig:supermomentum}, we plot the $L^{2}$ norm of the $\ell\geq2$ modes of the Moreschi supermomentum for the various CCE waveforms and the extrapolated waveform, both with and without the memory correction, for the equal mass non-spinning system. We see from this plot that before the $J_{\mathcal{E}}$ contribution is added, i.e., the memory correction, the extrapolated waveform appears to be much closer to the super rest frame than the CCE waveforms. However, when the expected time evolution is included, the final Bondi frame of the extrapolated waveform is closer to the frames of the CCE waveforms from the three largest world tube radii. Consequently, we observe that correcting the extrapolated waveforms with displacement memory not only decreases the violation of the BMS balance laws and thus the accuracy of the waveforms, but also makes important changes to the Bondi frame. Were one to try to apply a ``super rest correction,'' similar to that of a center-of-mass correction~\cite{PhysRevD.100.124010}, it is critically important to use memory-corrected extrapolated strain waveforms, rather than the raw extrapolated strain waveforms in the SXS and other waveform catalogs.

%%%%%%%%%%%%%%%%%%%%%%%%%%%%%%%%%%%%%%%%%%%%%%%%%%%%%%%%%%%
\section{Conclusion}
%%%%%%%%%%%%%%%%%%%%%%%%%%%%%%%%%%%%%%%%%%%%%%%%%%%%%%%%%%%
\label{sec:conclusion}

Gravitational memory is a unique physical observable that occupies the low frequency range and will most likely be measured by a future gravitational wave detector, such as LIGO Voyager, the Einstein Telescope, or LISA. Consequently, it is imperative that the waveforms that are produced by numerical relativity include memory so that when such an effect is detected we can check for any discrepancies with general relativity. At present, however, while the waveforms produced by CCE exhibit memory, the many waveforms that are made publicly available in the SXS catalog~\cite{SXSCatalog, Boyle_2019} and others~\cite{Jani_2016,Healy_2017} do not. Throughout this paper, we have demonstrated that the SXS catalog's extrapolated waveforms can be corrected by adding the contribution to the displacement memory that is sourced by the system's energy flux.

We started by checking that the oscillatory $J_{\Psi}$ contribution to the strain is indeed representative of the extrapolated strain waveform, as previously hypothesized. We then examined the BMS balance law violation both before and after applying our memory correction and found that adding the energy flux contribution improves the overall violation by roughly four orders of magnitude for 13 numerically evolved BBH systems that span a wide range of parameter space. After this, we noted that the main source of the remaining violation is from the time derivative of the spin memory contribution, which for an unknown reason is non-zero but is also not what is expected according to the BMS balance laws or by comparing to CCE waveforms. Finally, we showed that, besides satisfying the BMS balance laws, including the expected displacement memory also allows one to make a more correct measurement of the underlying system's BMS frame, which will prove vital for mapping waveforms to the super rest frame and making sure that waveforms computed by different methods are in the same frame.

\section*{Acknowledgments}

We would like to thank Dennis Pollney for sharing data that was used in the early stages of this project. Computations were performed with the High Performance Computing Center and on the Wheeler cluster at Caltech, which is supported by the Sherman Fairchild Foundation and by Caltech. This
work was supported in part by the Sherman Fairchild Foundation and by
NSF Grants No. PHY-2011961, No. PHY-2011968, and No. OAC-1931266 at Caltech, 
NSF Grants No. PHY-1912081 and No. OAC- 1931280 at Cornell, and NSF Grant No. PHY-1806356, Grant No. UN2017-92945 from the Urania Stott Fund of the Pittsburgh Foundation, and the Eberly research funds of Penn State at Penn State.

\begin{table*}
	\label{tab:runs}
	\centering
	\renewcommand{\arraystretch}{1.2}
	\begin{tabular}{@{}l@{\hspace*{7mm}}c@{\hspace*{7mm}}c@{}c@{}c@{}c@{\hspace*{7mm}}c@{}c@{}c@{}c@{}}
		\Xhline{3\arrayrulewidth}
		Name & $q$ & $\chi_{A}$:\, & $(\hat{x},\,$ & $\hat{y},\,$ & $\hat{z})$ & $\chi_{B}$:\, & $(\hat{x},\,$ & $\hat{y},\,$ & $\hat{z})$\\
		\hline
		\texttt{q1\_nospin}               & $1.0$ & & $(0,\,$ & $0,\,$ &  $0)$ & & $(0,\,$ & $0,\,$ & $0)$ \\
		\texttt{q1\_aligned\_chi0\_2}     & $1.0$ & & $(0,\,$ & $0,\,$ & $0.2)$ & & $(0,\,$ & $0,\,$ & $0.2)$ \\
		\texttt{q1\_aligned\_chi0\_4}     & $1.0$ & & $(0,\,$ & $0,\,$ & $0.4)$ & & $(0,\,$ & $0,\,$ & $0.4)$ \\
		\texttt{q1\_aligned\_chi0\_6}     & $1.0$ & & $(0,\,$ & $0,\,$ & $0.6)$ & & $(0,\,$ & $0,\,$ & $0.6)$ \\
		\texttt{q1\_antialigned\_chi0\_2} & $1.0$ & & $(0,\,$ & $0,\,$ & $0.2)$ & & $(0,\,$ & $0,\,$ & $-0.2)$ \\
		\texttt{q1\_antialigned\_chi0\_4} & $1.0$ & & $(0,\,$ & $0,\,$ & $0.4)$ & & $(0,\,$ & $0,\,$ & $-0.4)$ \\
		\texttt{q1\_antialigned\_chi0\_6} & $1.0$ & & $(0,\,$ & $0,\,$ & $0.6)$ & & $(0,\,$ & $0,\,$ & $-0.6)$ \\
		\texttt{q1\_precessing}           & $1.0$ & & $(0.487,\,$ & $0.125,\,$ & $-0.327)$ & & $(-0.190,\,$ & $0.051,\,$ & $-0.227)$ \\
		\texttt{q1\_superkick}            & $1.0$ & & $(0.6,\,$ & $0,\,$ & $0)$ & & $(-0.6,\,$ & $0,\,$ & $0)$ \\
		\texttt{q4\_nospin}               & $4.0$ & & $(0,\,$ & $0,\,$ & $0)$ & & $(0,\,$ & $0,\,$ & $0)$ \\
		\texttt{q4\_aligned\_chi0\_4}     & $4.0$ & & $(0,\,$ & $0,\,$ & $0.4)$ & & $(0,\,$ & $0,\,$ & $0.4)$ \\
		\texttt{q4\_antialigned\_chi0\_4} & $4.0$ & & $(0,\,$ & $0,\,$ & $0.4)$ & & $(0,\,$ & $0,\,$ & $-0.4)$ \\
		\texttt{q4\_precessing}           & $4.0$ & & $(0.487,\,$ & $0.125,\,$ & $-0.327)$ & & $(-0.190,\,$ & $0.051,\,$ & $-0.227)$ \\
		\Xhline{3\arrayrulewidth}
	\end{tabular}
	\caption{Parameters of the BBH mergers used in our results. The mass ratio is $q=M_A/M_B$, and the initial dimensionless spins of the two black holes are $\chi_A$ and $\chi_B$.}
\end{table*}

\newpage

%%%%%%%%%%%%%%%%%%%%%%%%%%%%%%%%%%%%%%%%%%%%%%%%%%%%%%%%%%%
%% BIBLIOGRAPHY
%%%%%%%%%%%%%%%%%%%%%%%%%%%%%%%%%%%%%%%%%%%%%%%%%%%%%%%%%%%

\bibliographystyle{apsrev4-1}
\bibliography{bibliography}

\end{document}